\documentclass[journal]{IEEEtran}
\usepackage{amsmath,amsfonts}
\usepackage{array}
\usepackage[caption=false,font=normalsize,labelfont=rm,textfont=rm]{subfig}
\usepackage{textcomp}
\usepackage{stfloats}
\usepackage{url}
\usepackage{verbatim}
\usepackage{graphicx}
\usepackage{cite}
\usepackage{balance}
\usepackage{subcaption}
\usepackage{bm}
\usepackage{orcidlink}
\usepackage[linesnumbered,ruled,vlined]{algorithm2e}
\begin{document}
%
\title{Bio-inspired Integrated Networking and Control for Large-Scale Swarm: A Hierarchical Co-design}


\author{Huan Lin~\orcidlink{0009-0002-7217-6139}, Dakai Liu, Lianghui Ding~\orcidlink{0000-0002-3231-3613 },~\IEEEmembership{Member,~IEEE}, Lin Wang~\orcidlink{0000-0002-8374-8473}, Feng Yang~\orcidlink{0000-0002-6350-5765}
\thanks{This paper was supported in part by Shanghai Key Laboratory Funding under Grant STCSM15DZ2270400, and in part by the Program for Professor of Special Appointment (Eastern Scholar) at Shanghai Institutions of Higher Learning. (\textit{Corresponding author: Lianghui Ding.})}
\thanks{Huan Lin, Chenguang Zhu, and Lianghui Ding are with the Institute of Image Communication and Network Engineering, School of Integrated Circuits, Shanghai Jiao Tong University, Shanghai 200240, China. Email: lhzt715@sjtu.edu.cn; robert\_liu@sjtu.edu.cn; lhding@sjtu.edu.cn.}
\thanks{Lin Wang is with the State Key Laboratory of Submarine Geoscience, School of Automation and Intelligent Sensing, Shanghai Jiao Tong University, Shanghai 200240, China. Email: wanglin@sjtu.edu.cn.}
\thanks{Feng Yang is with the Institute of Wireless Communication Technologies, School of Integrated Circuits, Shanghai Jiao Tong University, Shanghai 200240, China. Email: yangfeng@sjtu.edu.cn.}
}

%


\maketitle
\begin{abstract}
Unmanned aerial vehicle (UAV) swarms encounter the challenge of high overhead due to both network management and formation control requirements.  In this paper, we propose a Bio-inspired Integrated Networking and Control (BINC) scheme, enabling efficient formation management for swarms comprising thousands of UAVs. The scheme forms a two-layer hierarchical structure, where network clusters and formations share the same groups so that cross-cluster control is eliminated. For networking, we design a fused routing message together with control information to reduce overhead, and limit clusters' size to local two-hop topologies for fast command transmission. For controlling, we develop a hybrid bio-inspired control approach, including a pigeon-like leader-follower algorithm within formations under the consideration of cluster topology maintenance, and a starling-like algorithm among formations that helps to improve the ability of obstacle avoidance. 
We establish a simulation platform for UAV swarms with over 1000 nodes, and experimental results show that the proposed BINC scheme can achieve highly maneuverable swarm formation marching with significant reduction on communication overhead.
\end{abstract}


\section{Introduction}
\IEEEPARstart{U}nmanned aerial vehicles (UAVs) are nowadays applied in various military and civilian tasks, such as disaster relief, precision agriculture, monitoring \cite{mozaffari2019tutorial, adao2017hyperspectral, hu2020energy}, etc. Meanwhile, to overcome the limited capability of single UAVs, the development of UAV swarms consisting of hundreds or even thousands of UAVs enables the expanded applications on enhancing coverage, flexibility, resilience, and multitasking capabilities \cite{skorobogatov2020multiple, sun2017collision, demd, shakhatreh2019unmanned}. The UAV swarm forms a communication network to transmit task and control information. However, the increasing swarm scale brings an exponential growth of communication overhead for network management and formation control, leading to significant theoretical and engineering challenges\cite{9214446}. Therefore, both networking and controlling overhead need to be reduced in large-scale swarms.

The network management overhead is mainly caused by the message flood due to the dynamic swarm topology. Proactive routing protocols for mobile ad-hoc networks can be divided into flat and hierarchical routing. The flat routing protocols \cite{olsr, clsr, guizani2012new, kadadha2018cluster} utilize the multi-point relay (MPR) mechanism to limit the flooding within only critical nodes. Whereas, these flat routing protocols still need to transfer information of all nodes, leading to lengthy messages in the swarm network. The hierarchical routing protocols\cite{holsr, mao2017eholsr}, on the contrary, establish a multi-layer clustered architecture. By isolating the flooding messages between clusters, these protocols dramatically decrease the number of hops for message transfer and reduce the message length and networking overhead. 

On the other hand, the control overhead is inevitable, as the UAVs need to frequently transmit information such as position, speed, and obstacles to ensure safety during the flight process. The early studies primarily utilize centralized control algorithms \cite{tan1996virtual, lewis1997high, consolini2008leader, pigeon} for formations with multiple UAVs. The centralized implementations achieve stable control via high-performance control centers, but lack scalability to growing swarm sizes and varying environments. Subsequently, distributed control algorithms \cite{reynolds1987flocks, balch1998behavior, wu2019formation} were developed and later extended for large-scale swarm control \cite{starling, long2020comprehensive, li2022large, bu2024advancement}. The distributed intelligence of UAVs enables maneuverable and flexible swarm control with capabilities of collision avoidance, obstacle navigation, and path planning.

In contrast to isolated approaches for networking or control, swarm intelligence offers a promising paradigm: integrating communication protocols with formation control algorithms to achieve high-performance swarm management. For instance, approaches such as \cite{pradittasnee2016efficient} leverage optimized routing to expedite critical control commands, while techniques in \cite{yang2021neural, liu2022task} enforce trajectory constraints to preserve network connectivity. However, these studies primarily focus on unidirectional optimization such that enhancing control via networking or maintaining network through control, while largely overlooking the inherent bidirectional constraints and synergies between the two domains.

To achieve deeply coupled co-design between networking and control, three challenges are investigated in designing our scheme. First, hierarchical structure design demands careful trade-offs, as control algorithms may require information from inter-cluster members that is  transmitted inefficiently.
Second, scalability remains a fundamental barrier, since the message explosion resulted by both networking and controlling requirements in swarms scaling to thousands of UAVs would occupy all available capacity. 
The third challenge lies on the control decisions without topology maintenance awareness, which can amplify reconstruction frequency and undermines system robustness.

In this paper, we propose a Bio-inspired Integrated Networking and Control (BINC) scheme, aiming to achieve high maneuverability with low communication overhead for large-scale swarm. 
The scheme refers the hierarchical architecture for networking according to \cite{holsr}, but further unifies formations and clusters under the same structure, hence avoiding control command transmission between inter-cluster members.
We improve the hierarchical network routing protocol for better transmitting control information. In particular, BINC scheme incorporates control information into routing messages, and restricts the cluster size to the two-hop neighbors of the cluster head. The fused message design can reduce the message overhead, while the limited cluster scale ensures fast topology awareness and command dispatch.
Meanwhile, we develop a hybrid bio-inspired control algorithm to fit to the hierarchical architecture. The pigeon-like leader-follower algorithm within a formation enables dense and stable formation flight in dynamic environments. The starling-like algorithm among formations achieves distributed swarm marching, improving obstacle and collision avoidance capabilities. This hybrid control algorithm effectively maintains the original clustered topology and thus reduce the network maintenance overhead.
We also establish a simulation platform for large-scale UAV swarms, including both the networking protocol stack and the formation control model. Experimental results show that the proposed BINC scheme can achieve highly maneuverable swarm control with an average of 70.5\% reduction of communication overhead. Our contributions are summarized as follows.
\begin{enumerate}
    \item We propose a novel integrated network managing and motion controlling scheme, BINC. This scheme achieves, for the first time, the simulation of distributed control in an ad hoc network of 1,000 nodes, expanding the application boundaries of swarm networks.
    \item We unify the network protocol and control algorithm based on same hierarchical architecture and message structure, mitigating the misalignment between clusters and formations while reducing message overhead. This design turns the mutual constraints between networking and controlling into mutual promotion.
    \item We develop a hybrid bio-inspired control approach for maneuverable swarm under the hierarchical structure. The intra-formation pigeon-like algorithm enables stable node-level control for cluster maintenance, while the inter-formation starling-like algorithm provides highly flexible swarm-level control.
\end{enumerate}

The remainder of the paper is organized as follows. Section \uppercase\expandafter{\romannumeral2} reviews the related work. Section \uppercase\expandafter{\romannumeral3} presents the system model of the UAV swarm. Section \uppercase\expandafter{\romannumeral4} describes the implementation details of the proposed scheme, and the performance is evaluated in Section \uppercase\expandafter{\romannumeral5}. Finally, the conclusion is summarized in Section \uppercase\expandafter{\romannumeral6}.

\section{Related work}
Communication overheads in large-scale UAV swarms consist of network management and formation control overhead. The former aims to maintain reliable routing to transfer information, while the latter provides the necessary data to accomplish a safe flight mission. Therefore, the routing protocols are the foundation for the control algorithms to exchange their commands, and location changes of UAVs directed by the control algorithms will also affect the network to update the route.

\subsection{Network Management Protocols}
The UAV swarm network is a typical mobile ad-hoc network (MANET), where nodes and links are dynamic. Network routing protocols for MANET are mainly proactive, as each node needs to maintain a local routing table to achieve self-organized networking. The OLSR protocol in \cite{olsr}, the most famous routing algorithm for MANET, designed the HELLO message for neighbor discovery and the TC message for topology delivery. Meanwhile, the multi-point relay (MPR) mechanism in OLSR minimized the redundancy of topology information flooding, establishing a foundation for reducing routing overhead in large-scale networks. Subsequently, numerous research achievements have been made in improving the OLSR. Some research efforts aimed to reduce the routing overhead by optimizing the flooding frequency, but showed limited effectiveness \cite{guizani2012new}. Other approaches proposed to optimize the MPR mechanism \cite{rivoirard2017multipoint,song2021pf}. Protocols such as CLSR\cite{clsr} and C-OLSR\cite{kadadha2018cluster} combined the MPR selection with clustering structure, reducing the number of flooding messages at the cost of sacrificing optimal routing.

However, routing algorithms mentioned above encounter a key challenge, that is, their flat routing structure leads to tremendous storage and computation requirements in networks with thousands of nodes. A feasible approach is to design a hierarchical architecture that can reduce the routing overhead by an order of magnitude. For example, Villasenor \textit{et al.} \cite{holsr} implemented the HOLSR that forms a multi-layer network based on node capability and link capacity, and Mao \textit{et al.} \cite{mao2017eholsr} utilized k-means clustering to optimize the layer segmentation. The hierarchical structure results in OLSR having to perform a reduced amount of routing computations as the local movement of member nodes is now handled locally, not affecting other parts of the network.
Therefore, this paper enhances the routing scalability by altering the hierarchical network architecture of large-scale UAV swarms.

\subsection{Motion Control Algorithms}
Control algorithms for UAV formations and swarms can be divided into centralized and distributed ones. The centralized algorithms relied on control centers in each formation to determine the positions or trajectories. Early studies \cite{tan1996virtual,lewis1997high} designed a virtual architecture to achieve precise formation control, but lacked scalability when the number of UAVs changed. On the contrary, the leader-follower approach can handle the varying formation scales, as the leader calculates the relative positions of each follower through predefined rules \cite{consolini2008leader}. The integration of feedback-linear control, model-predictive control, back-stepping techniques, multi-leader strategies, and virtual leader approaches further enhances the control effectiveness of the formation and enriches its functionality \cite{turpin2012trajectory,he2018feedback,ali2021multi}. However, when the swarm scale reaches hundreds or thousands, control centers suffer from severe latency of information aggregation and command transfer, which undermines the effectiveness of swarm control.

Distributed control algorithms aim to utilize the UAVs' intelligence to achieve collision avoidance, obstacle navigation, and path planning. Therefore, the UAVs need to broadcast their data for collaborative flight, e.g., locations, velocities, directions, accelerations, etc. Reynolds \textit{et al.} \cite{reynolds1987flocks} introduced the Boids algorithm with three behavior principles: separation, alignment, and cohesion. Then, several behavior-based algorithms improve the Boids algorithm to adapt to large-scale swarms \cite{long2020comprehensive, li2022large, bu2024advancement}, but these approaches require meticulous design of the principles. Artificial intelligence-based methods are also applied in formation control currently. For instance, Liu \textit{et al.} \cite{liu2017neural} introduced neural networks for path planning, while Li \textit{et al.} \cite{li2019autonomous} developed the deep Q-learning network to minimize energy consumption and obstacle avoidance. Whereas, artificial intelligence-based methods require a large amount of data for training, which poses high demands on the computing and storage capabilities.

In this paper, we focus on a valuable category of swarm control approaches — the bio-inspired algorithms. The Boids algorithm is the most representative one, which is inspired by studying the patterns of bird flocks. Later research explored the behaviors of different animal flocks. For example, Nagy \textit{et al.} \cite{pigeon} revealed the leader-follower relationship within the pigeon flocks for formation control, and Williams \textit{et al.} \cite{williams2015pigeons} demonstrated that the hierarchies and social structures enable the pigeon flocks to adapt to external changes. Hamed \textit{et al.} \cite{hamed2022hunting} proposed a collaborative hunting strategy based on the wolf swarm algorithm to hunt dynamic targets. For swarm control, Young \textit{et al.} \cite{starling} modeled the restricted vision of starling flocks. Studies on fish schools \cite{brown2011fish} and bees \cite{gao2019overview} also highlighted the fundamental behaviors of alignment, cohesion, and collision avoidance. The leverage of the internal behaviors observed in biological systems can significantly enhance the capacity of autonomous formation control \cite{li4941922formation}.

\subsection{Integrated Networking and Control Co-design}
In terms of research on the integrated design of formation control and network architecture, most existing studies model it as an optimization problem. For example, Pradittasnee \textit{et al.}  \cite{pradittasnee2016efficient} designed the optimal routing to accelerate the transmission of emergency control messages so that the communication delays meet the formation control requirements. Yang \textit{et al.} \cite{yang2021neural} integrated neural networks and leader-follower approaches to achieve formation control, collision avoidance, and connectivity maintenance. Liu \textit{et al.} \cite{liu2022task} achieved higher coding rate thresholds and lower system instability probabilities by optimizing formation control parameters. Chang \textit{et al.} \cite{chang2021autonomous} transformed the control convergence rate into communication reliability constraints to optimize communication power and device activation probability, while Zhao \textit{et al.} \cite{zhao2023integrated} optimized the effect of formation control by adjusting the allocation algorithm of sub-carriers and the variation mode of communication rate. Overall, the current research on the integration of formation control and network architecture operates at a relatively low level, and cannot address the adaptability issues in large-scale swarm scenarios.

This paper aims to explore a more efficient integrated networking and control co-design to adapt to swarm marching missions with over 1000 UAVs. By dividing the entire swarm into two layers, we consider a consistent hierarchical structure, i.e., clusters and formations share the same members. The network routing provides fast control command transfer within the formations, while the formation control helps to maintain a stable cluster topology, thereby achieving efficient and maneuverable swarm control with lower communication overheads.

\section{System Model}
This section first constructs the network model to represent the UAV swarm, including the clustered topology and communication ability, and then introduces the motion model of individual UAVs and formations.

\subsection{Communication Channel and Network Model}
Assuming that the swarm is composed of $N$ independent and identical UAVs, where all UAVs are labeled by a global index $\mathcal{I}=\{1,2,..., N\}$. For most missions, the UAVs mainly form a network at equal altitudes, hence, we only consider two-dimensional coordinates of UAVs' locations. Let location vector $\bm{p}_i=[x_i, y_i]^\top\in \mathbb{R}^2$ denote the coordinate of the UAV $u_i$, where $x_i$ and $y_i$ are the coordinate components of the $X$ and $Y$ axes, respectively.

Based on a cluster formulation algorithm, the swarm network is divided into $M$ groups as $\mathcal{G}_{all}=\{\mathcal{G}_1, \mathcal{G}_2,..., \mathcal{G}_M\}$, where $\mathcal{G}_m=\{m_1, m_2,..., m_k\}$ denotes the index of UAVs in the $m$-th group. In particular, one of the UAVs in the group is selected as the cluster head $h_m$, so that we can form the head index as $\mathcal{I}_H=\{h_1,h_2,..., h_M\}$, which represents the second layer of the hierarchical structure.

Inspired by \cite{holsr}, we consider that each UAV is equipped with two distinct communication channels: a long-distance channel and a short-distance channel at different frequencies. The long-distance channel is applied  between cluster heads with a transmission range $D_{tr}$. In contrast, the short-distance channel is used for communication with neighboring UAVs, offering reduced energy consumption with a transmission range $d_{tr}$.

We first model the transmission on the short-distance communication channel. Without loss of generality, we consider a predominantly line-of-sight communication link between UAVs in the air. According to Friis transmission formula, the received signal power of $u_j$ from $u_i$ is
\begin{align}
    P_{ij}=P_0G_{tx}G_{rx}L(d_{ij})|h_0|^2,
\end{align}
where $P_0$ is the transmitting power, $G_{tx},G_{rx}$ are the constant gains of the transmitting and receiving antennas, respectively. $L(d_{ij})=\left(\lambda_c/{4\pi d_{ij}}\right)^2$ is the large-scale fading, where $\lambda_c$ is the wavelength, $d_{ij}=\|\bm{p}_i-\bm{p}_j\|$ is the Eulerian distance between the UAV $u_i$ and $u_j$. The $|h_0|^2$ is the small-scale fading that follows the Gamma distribution. Denoting the receiver sensitivity $\tau$, condition of UAVs $u_i$ and $u_j$ that can form a communication link follows
\begin{align}
    P_0G_{tx}G_{rx}L(d_{ij})|h_0|^2\leq\tau.
    \label{dtr}
\end{align}
We can then calculate the maximum transmission range $d_{tr}$ of the short-distance channel by solving (\ref{dtr}) as an equation. UAVs can establish a communication link through the short-distance channel when $\|\bm{d}_{ij}\|\leq d_{tr}$. In this case, the UAVs $u_i$ and $u_j$ are so-called $1$-hop neighbors, and the number of $1$-hop neighbors denotes the connectivity degree $c_i$ of node $u_i$.

Similarly, with given frequency and transmitting power on the long-distance channel, we can also calculate the maximum transmission range $D_{tr}$ according to (\ref{dtr}). Therefore, as considered in our network model, the cluster heads $u_{h_i}$ and $u_{h_i}$ can establish a communication link through the long-distance channel if and only if $\|\bm{d}_{h_ih_j}\|\leq D_{tr}$.

\begin{figure*}[!t]
\centerline{\includegraphics[width=\linewidth]{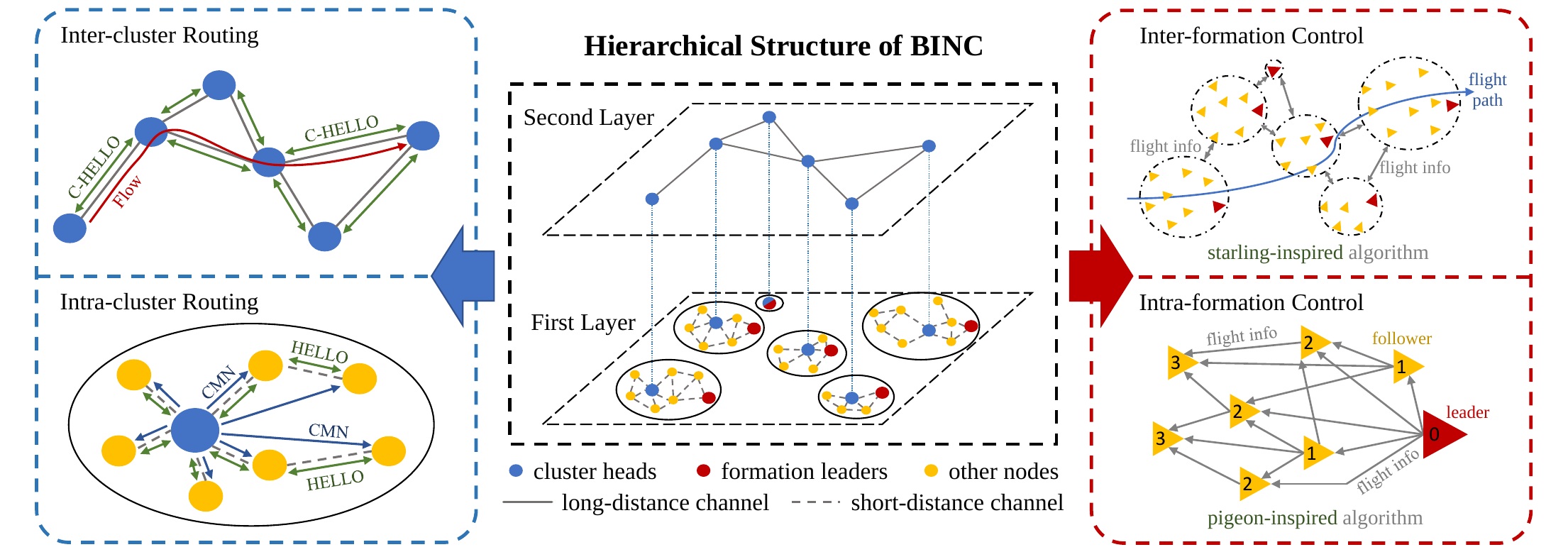}}
\captionsetup{justification=raggedright, singlelinecheck=false}
\caption{Overall framework of the proposed BINC scheme.}
\label{framework}
\end{figure*}

\subsection{Motion Model}
Consider the motion model on the node level, in this paper, we regard each UAV as a mass point to simplify the flight control. The control equation governing the dynamics of the UAV $u_i$ is depicted as
\begin{equation}
    \begin{cases}
        \begin{array}{l}
            \Delta\bm{p}_i=\bm{v}_i\cdot\Delta t\\
            \Delta\bm{v}_i=\bm{f}_i\cdot\Delta t\\ 
         \end{array} \\
    \end{cases}
    \label{e1}
\end{equation}
where $\bm{v}_i, \bm{f}_i\in \mathbb{R}^2$ denote the velocity and the applied force of $u_i$, respectively, and $\Delta t$ is the time step for control. Therefore, the node-level control algorithm aims to calculate $\bm{f}_i$ for each UAV $u_i$.

Notice that the swarm has a hierarchical structure; UAVs of the same group form a control formation, and each formation is considered an entity for inter-formation control. Therefore, we need to further determine the motion model of formations. For the $m$-th formation with index $\mathcal{G}_m$, its head $u_{h_m}$ calculates the formation center as 
\begin{align}
    \bm{P}_m = \frac{1}{\left| \mathcal{G}_m \right|}\mathop{\sum}\limits_{k\in\mathcal{G}_m}\bm{p}_k.
\end{align}
The formation is then formulated as a circle with center $\bm{P}_m$, and the radius is calculated by the maximum distance of $\bm{P}_m$ to each UAVs in the formation, i.e.,
\begin{align}
    R_m=\mathop{\text{max}}\limits_{k\in\mathcal{G}_m}\|\bm{P}_m-\bm{p}_k\|.
    \label{radius}
\end{align}

Since the follower nodes in a formation are controlled by their leader UAV, where we assume $L_m$ denotes the leader node of formation $\mathcal{G}_m$, the formation motion model can be simplified as the motion model of the leader. Similar to (\ref{e1}), the control equation governing the dynamics of the formation $\mathcal{G}_m$ is depicted as
\begin{equation}
    \begin{cases}
        \begin{array}{l}
            \Delta\bm{P}_m=\bm{v}_{L_m}=\bm{V}_m\\
            \Delta\bm{V}_m=\bm{F}_m\\ 
         \end{array} \\
    \end{cases}
    \label{e2}
\end{equation}
where $\bm{v}_{L_m}$ is the velocity of the leader UAV of formation $\mathcal{G}_m$, and $\bm{V}_m, \bm{F}_m\in \mathbb{R}^2$ are the velocity and the applied force, respectively. In particular, the velocity of the leader UAV is set to be equal to the formation velocity, and thus the inter-formation control algorithm aims to determine $\bm{F}_m$ for each formation.

\section{Bio-inspired Integrated Networking and Control scheme}
This section first introduces the overall architecture of the proposed BINC scheme. Then the second subsection details the network management and routing implementation. The bio-inspired intra and inter-formation control algorithms are presented in the third and fourth subsections, respectively.

\subsection{Overall Architecture of BINC}
As shown in Fig.\ref{framework}, nodes in the swarm network are divided into two layers with several groups based on a multi-round clustering algorithm. Each group $\mathcal{G}_m$ has a routing head and a formation leader. The nodes in the first layer communicate with their neighbors through the short-distance channel, and the heads in the second layer utilize the long-distance channel for inter-cluster information transmission. This hierarchical structure is shared by both network management and swarm control.

The network management is illustrated in the left part of Fig.\ref {framework}, which consists of the intra-cluster and inter-cluster routing. Similar to the HOLSR protocol \cite{holsr}, the routing message design contains the HELLO messages for neighbor discovery and the TC messages for topology awareness. The HTC messages that use the MPR mechanism to broadcast the cluster topology to other cluster heads are also applied in our routing mechanism. Additionally, we design two novel messages, the Cluster-Hello (C-HELLO) messages and the Cluster-Member-Notification (CMN) messages, for head selection and cluster maintenance. 

The swarm control is shown in the right part of Fig.\ref{framework}. For intra-formation control, we develop a pigeon-inspired leader-follower algorithm that transfers the flight information based on the social level of each UAV. Meanwhile, as each group is considered as a formation, we apply a starling-inspired algorithm to achieve inter-formation control according to the flight information from the local area. All the flight information is embedded into the routing messages to decrease the frequency of control command transmission. Therefore, a swarm with the BINC scheme can achieve maneuverable turning and obstacle avoidance with low communication overhead.


\subsection{Networking and Routing in BINC}
The network management in the BINC scheme includes two phases: network establishment and network maintenance. We first detail the fused message design in BINC as the foundation of network management. Then, we develop a multi-round clustering algorithm to establish the hierarchical swarm network. Finally, the network maintenance and routing mechanism are introduced to achieve reliable communication under dynamic scenarios.

\subsubsection{Fused message design}
The messages in BINC are designed to contain both routing and flight information. The HELLO messages are used for neighbor interaction inside clusters via the short-range communication channel. Each node generates its own HELLO message that mainly contains the following information:
\begin{itemize}
    \item The connectivity degree and ID of this node.
    \item The position and velocity of this node.
    \item The Head ID and Leader ID of this node.
    \item The identification sequence number for velocity loop suppression.
    \item The connectivity degrees, IDs, and positions of its one-hop neighbors.
\end{itemize}

The C-HELLO messages are designed for neighbor interaction between cluster heads through the long-distance channel. Each head node generates a C-HELLO message that mainly contains the cluster information as:
\begin{itemize}
    \item The center position of this group.
    \item The equivalent radius and velocity of this group.
    \item The Head ID and Leader ID of this group.
    \item The identification sequence number for velocity loop suppression.
    \item The Head IDs, positions, radii, and velocities of its one-hop neighboring groups.
\end{itemize}

The CMN messages are developed for cluster head and member management within a cluster. The head node generates the CMN messages and transfers them to its member nodes through the short-distance channel, which includes the information as follows:
\begin{itemize}
    \item The force of this group.
    \item The Head ID of this group.
    \item The Member IDs and social ranks of member nodes in this group.
\end{itemize}

\begin{figure}[t]
    \centering
    \includegraphics[width=\linewidth]{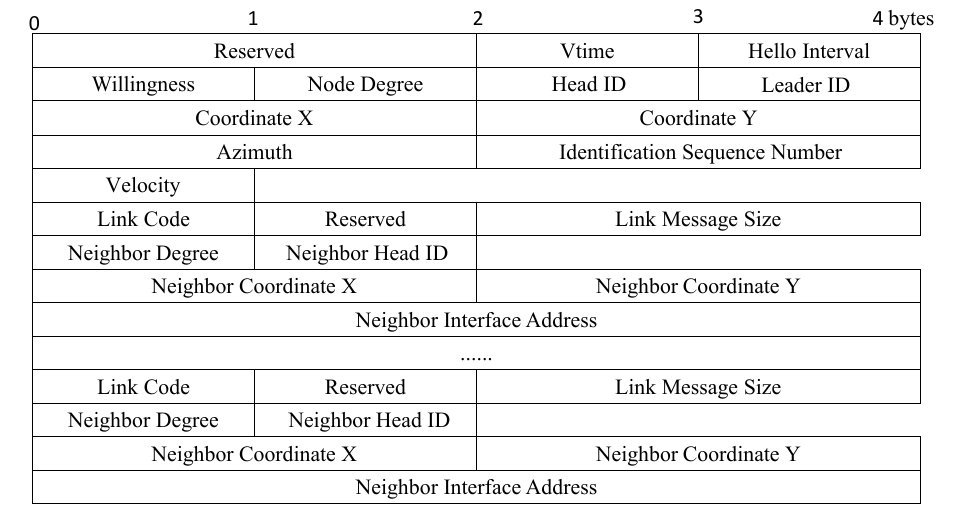}
    \captionsetup{justification=raggedright, singlelinecheck=false}
    \caption{The packet format of HELLO messages.}
    \label{hello}
\end{figure}
\begin{figure}[t]
    \centering
    \includegraphics[width=\linewidth]{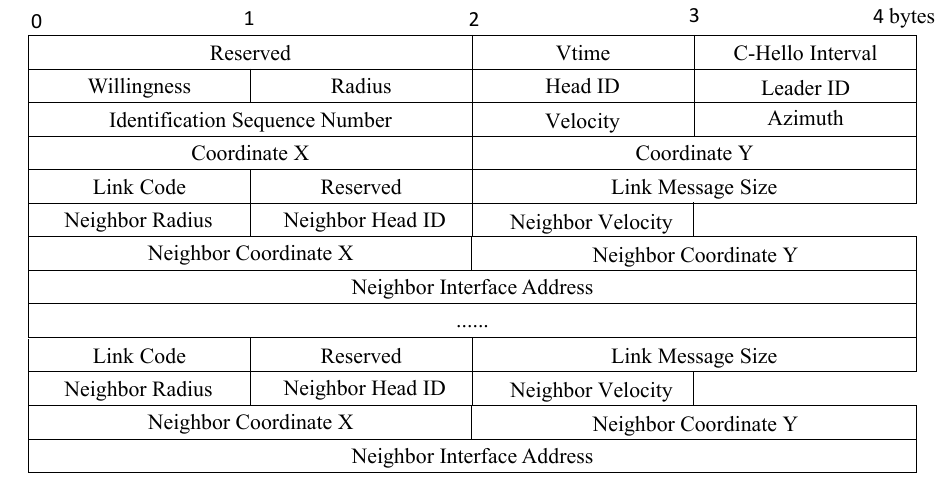}
    \captionsetup{justification=raggedright, singlelinecheck=false}
    \caption{The packet format of C-HELLO messages.}
    \label{chello}
\end{figure}
\begin{figure}[!t]
    \centering
    \vspace{-0.5em}
    \includegraphics[width=\linewidth]{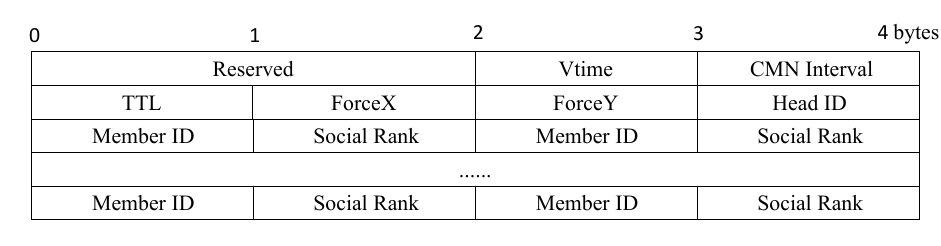}
    \captionsetup{justification=raggedright, singlelinecheck=false}
    \caption{The packet format of CMN messages.}
    \label{cmn-format}
\end{figure}

The packet formats for HELLO, C-HELLO, and CMN messages are illustrated in Fig.\ref{hello}, Fig.\ref{chello}, and Fig.\ref{cmn-format}, respectively.
For topology control (TC) messages, our BINC scheme utilizes the same design according to HOLSR \cite{holsr}, i.e., TC messages for intra-cluster routing and HTC messages for inter-cluster routing. Therefore, the routing messages can help to transmit basic control information, reducing extra overhead for generating control messages individually.

\subsubsection{Network clustering with size limitation}
According to the connectivity degree information from HELLO messages, nodes compare the connectivity degree information of their neighbors, and cluster heads are elected through multiple rounds of elections. To distinguish the clustering states, we determine two states of nodes as \textit{clustered} and \textit{unclustered}. The initial state of a node is \textit{unclustered}. When a node is judged as the maximum-degree node within the $2$-hop range, it will convert it into \textit{clustered} and transfer CMN messages to its $2$-hop neighbors. Once an \textit{unclustered} node receives the CMN message, it converts the state to \textit{clustered}. So far, the swarm network has completed the first round of clustering.  

However, the network after the first-round clustering still has several \textit{unclustered} nodes, as the selected head nodes cannot always cover the entire network topology. Therefore, we develop multiple external rounds of clustering to ensure that all nodes are determined as \textit{clustered}. We set a MAX-WAITING-TIME to trigger the external clustering rounds that each \textit{unclustered} node will exclude its \textit{clustered} neighbors and recalculate its $c_i$. Similarly, a node with the largest degree within a $2$-hop range becomes a cluster head and floods the CMN message to attract the unclustered nodes. After multiple rounds of clustering, the states of all nodes are changed to the \textit{clustered} state.

\subsubsection{Network maintenance principle}
The swarm movement enables a dynamic network topology, where clusters may face the scenarios of splitting, overlapping, and merging. These changes can lead to confusion in cluster management: some nodes lose CMN messages from cluster heads, while others may receive CMN messages from different cluster heads at the same time. To enhance the robustness of the clustered hierarchical structure, we designed several principles for network maintenance, as follows.

\begin{itemize}
    \item Cluster splitting: this occurs when members move away from the cluster. The members leaving the cluster wait for MAX-WAITING-TIME, and if no CMN message is received, the members trigger a head selection round.
    \item Cluster overlapping: this occurs when two UAV clusters intersect during formation movement and causes some members to receive both CMN messages from two heads. They compare its hop counts to two heads and choose the shorter one. If the hop counts are the same, the members stay put.
    \item Cluster merging: this occurs when two cluster heads become $2$-hop neighbors. Both get each other's CMN messages, so one must step down. Via HELLO messages, they compare the connectivity degree. The one with the lower degree joins the other cluster, and the members in this group may re-select the head according to MAX-WAITING-TIME.
\end{itemize}

\subsubsection{Network routing}
We separately introduce the routing mechanism for member nodes and head nodes.
Member nodes only operate on the intra-cluster channel. Each member node broadcasts its own HELLO messages and forwards the CMN messages. The MPR member nodes also flood the TC messages inside the cluster. Meanwhile, head nodes work on both intra- and inter-cluster channels. Each head node generates HELLO, TC, and CMN messages within the cluster, and transfers C-HELLO and HTC messages for inter-cluster interaction. 

For the routing process of transmitting a flow, if the destination node is within the cluster, the packet is directly forwarded to it. If the destination node is outside the cluster, the packet is sent to the head node by default. The head node then searches for the cluster where the destination node resides, forwards the packet to the target head node first, and the target head node finally forwards it to the destination node. Specifically, as the flight information is embedded into the routing messages, the BINC scheme requires no external flow for formation and swarm control, which significantly reduces the communication overhead.

\subsection{Intra-formation Control in BINC}
The intra-formation control aims to determine the force $\bm{f}_{m_i}$ for the UAV $u_{m_i}$ in the formation group $\mathcal{G}_m$. Inspired by the motion behavior of pigeon flocks \cite{pigeon}, we develop a pigeon-like leader-follower control algorithm to achieve reliable formation control in the first layer. In general, the force $\bm{f}_{m_i}$ is divided into two parts, i.e.,
\begin{align}
    \bm{f}_{m_i} = \bm{f}_{m_i}^N+\bm{f}_{m_i}^F,
    \label{f_mi}
\end{align}
\begin{table*}[!b]
\normalsize
\centering
\rule{\linewidth}{1pt}
\begin{equation}
\bm{f}_{m_im_j}^N =
\begin{cases}
    -\dfrac{\bm{d}_{m_im_j}}{\|\bm{d}_{m_im_j}\|}\left(\cos(\dfrac{\pi}{2}\cdot\dfrac{\|\bm{d}_{m_im_j}\|}{r_{rep}}) + A\right), & 0<\|\bm{d}_{m_im_j}\|\leq r_{rep}.\\
    0, & r_{rep}<\|\bm{d}_{m_im_j}\|\leq r_{al}.\\
    \dfrac{\bm{d}_{m_im_j}}{\|\bm{d}_{m_im_j}\|}\left(\cos(\dfrac{\pi}{2}\cdot\dfrac{r_{att}-\|\bm{d}_{m_im_j}\|}{r_{att}-r_{al}}) + A\right), & r_{al}<\|\bm{d}_{m_im_j}\|\leq r_{att}.
\end{cases}
\label{pneighbor}
\end{equation}
\end{table*}
where $\bm{f}_{m_i}^N$ is the resultant force of neighbor interactions in the group, and $\bm{f}_{m_i}^F$ is the following force caused by the leader-follower requirement. Then, we determine the force components for the follower nodes.

\begin{figure}[!t]
\vspace{-0.5em}
\centerline{\includegraphics[width=.96\linewidth]{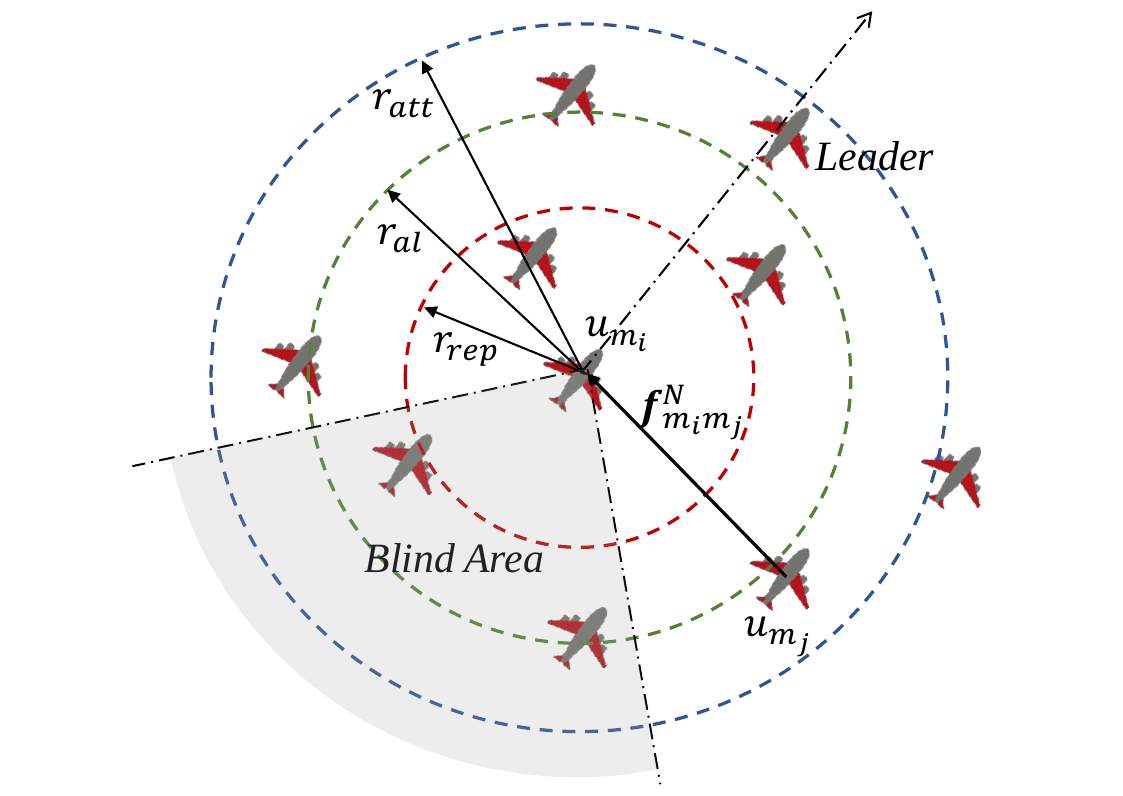}}
\captionsetup{justification=raggedright, singlelinecheck=false}
\caption{Neighbor interaction model of pigeon-like formation control.}
\vspace{-0.5em}
\label{pigeon-alg}
\end{figure}

\subsubsection{Neighbor-interaction force}
Bird flocks typically perform three motion behaviors: separation, alignment, and cohesion, which are determined by the distance between two birds. The neighbor interaction model of pigeon-like formation control is shown in Fig.\ref{pigeon-alg}. Each UAV divides its perception range into the Repel Area, Aline Area, and Attract Area with the area radii $r_{rep}$, $r_{al}$, and $r_{att}$. Meanwhile, due to the limited vision, the node only recognizes its neighbors outside the Blind Area.

Therefore, node $u_{m_i}$ forms a interactive neighbor set as $\mathcal{U}_{m_i}^N=\{u_{m_j}|\|\bm{d}_{m_im_j}\|\leq r_{att}, m_j\in\mathcal{G}_m ,m_j\notin Blind\ Area\}$, where $\bm{d}_{m_im_j}=\bm{p}_{m_j}-\bm{p}_{m_i}$ is the location difference between nodes $u_{m_i}$ and $u_{m_j}$. As we use short-range communication links for neighbor discovery in routing, $r_{att}$ is set smaller than $d_{tr}$ to quickly identify whether neighboring UAVs belong to the same group and perform corresponding actions based on neighbor types, achieving functions such as obstacle avoidance and following.

For each $u_{m_j}\in\mathcal{U}_{m_i}^N$, the neighbor will result in a neighbor-interaction force according to the distance $\|\bm{d}_{m_im_j}\|$. The force caused by neighbors is calculated by (\ref{pneighbor}), where $A$ is a constant positive coefficient to enhance the interaction force. Neighbors in the Attract Area have attractive effects, while neighbors in the Aline Area do not generate any force. Other neighbors in the Repel Area lead to repulsive actions. Therefore, the resultant interactive force of $u_{m_i}$ is determined as
\begin{align}
\bm{f}_{m_i}^N=\mathop{\sum}\limits_{u_{m_j}\in\mathcal{U}_{m_i}^N}\bm{f}_{m_im_j}^N.
\end{align}

\subsubsection{Following force}
In nature, pigeon flocks have a strict social hierarchy, where lower-level pigeons will follow higher-level pigeons. The structure of a pigeon flock formation is shown in the lower-right part of Fig.\ref{framework}, where each node has a corresponding social level and the flight information is transmitted from higher-level nodes to lower-level nodes. Therefore, the prerequisite for controlling the formation of pigeon-inspired flocks is the allocation of social levels. 

In the flight direction of the formation, we first select the node located at the forefront as the leader node $u_{l_m}$ of group $\mathcal{G}_m$ with social level $S_{l_m}=0$. For each follower $u_{l_m}$ in the formation, we assign social levels according to the number of hops between the leader and the follower, i.e., $S_{m_i}=\text{Hop}(u_{l_m},u_{l_m})$. Since the members in a group are all within the $2$-hop range of the cluster head, the social levels in a formation have an upper bound as $S_{max}=4$. To limit the following effects within a local range, we provide each follower node with a following node set as $\mathcal{N}_{m_i}^F\{u_{m_j}|0<S_{m_j}-S_{m_i}\leq 2, m_j\in\mathcal{G}_m\}$.

The leader-follower effect can be calculated by two components: position following and velocity alignment. The position following force aims to maintain the compactness of the pigeon-like formation, where the follower node is attracted by its superior nodes in its superior node set. Hence, we can calculate the position-following force of $u_{m_i}$ as
\begin{align}
\bm{f}_{m_i}^{p}=\mathop{\sum}\limits_{u_{m_j}\in\mathcal{U}_{m_i}^F}\frac{\bm{d}_{m_im_j}}{\|\bm{d}_{m_im_j}\|}.
\end{align}

In terms of velocity alignment, the follower node tends to have the same velocity as the leader node. The leader node determines its velocity and passes on the velocity through the social hierarchy mechanism. After receiving the velocity from neighboring nodes, the follower nodes with lower social levels set their velocities to the average velocity of their higher-level neighbors. Finally, the entire formation reaches a consistent velocity and flies along the predetermined trajectory. The velocity-alignment force is represented as
\begin{align}
\bm{f}_{m_i}^{v}=\frac{1}{|\mathcal{U}_{m_i}^F|}\mathop{\sum}\limits_{u_{m_j}\in\mathcal{U}_{m_i}^F}(\bm{v}_{m_j}-\bm{v}_{m_i}).
\end{align}

The following force of the follower node in each formation is then calculated as
\begin{align}
    \bm{f}_{m_i}^F=\bm{f}_{m_i}^p+\bm{f}_{m_i}^v.
\end{align}

Overall, the formation control of pigeon-like groups within the cluster adopts a distributed control approach to reduce the control complexity and communication overhead. To achieve orderly formation control and rapid velocity convergence, the head node periodically allocates the social level of nodes within the group through the CMN message. When the leader node is damaged or more suitable leaders appear, it can quickly replace the leader, enhancing the robustness of formation control.

\subsection{Inter-formation Control in BINC}
The inter-formation control aims to determine the force $\bm{F}_{m}$ for the formation group $\mathcal{G}_m$. We developed a starling-like algorithm to control the relative distance and velocity between formations, so that UAVs can avoid interference from other UAVs and coordinate the flight paths between groups. Similarly, we divided the applied force of a formation as
\begin{align}
    \bm{F}_{m}=\bm{F}_{m}^N+\bm{F}_{m}^P,
\end{align}
where $\bm{F}_{m}^N$ is the resultant interactive force caused by neighboring groups, and $\bm{F}_{m}^P$ is the force caused by the current pattern of the starling-like formation. Then, we determine the force components for each formation group.

\begin{figure}[!t]
\centerline{\includegraphics[width=.96\linewidth]{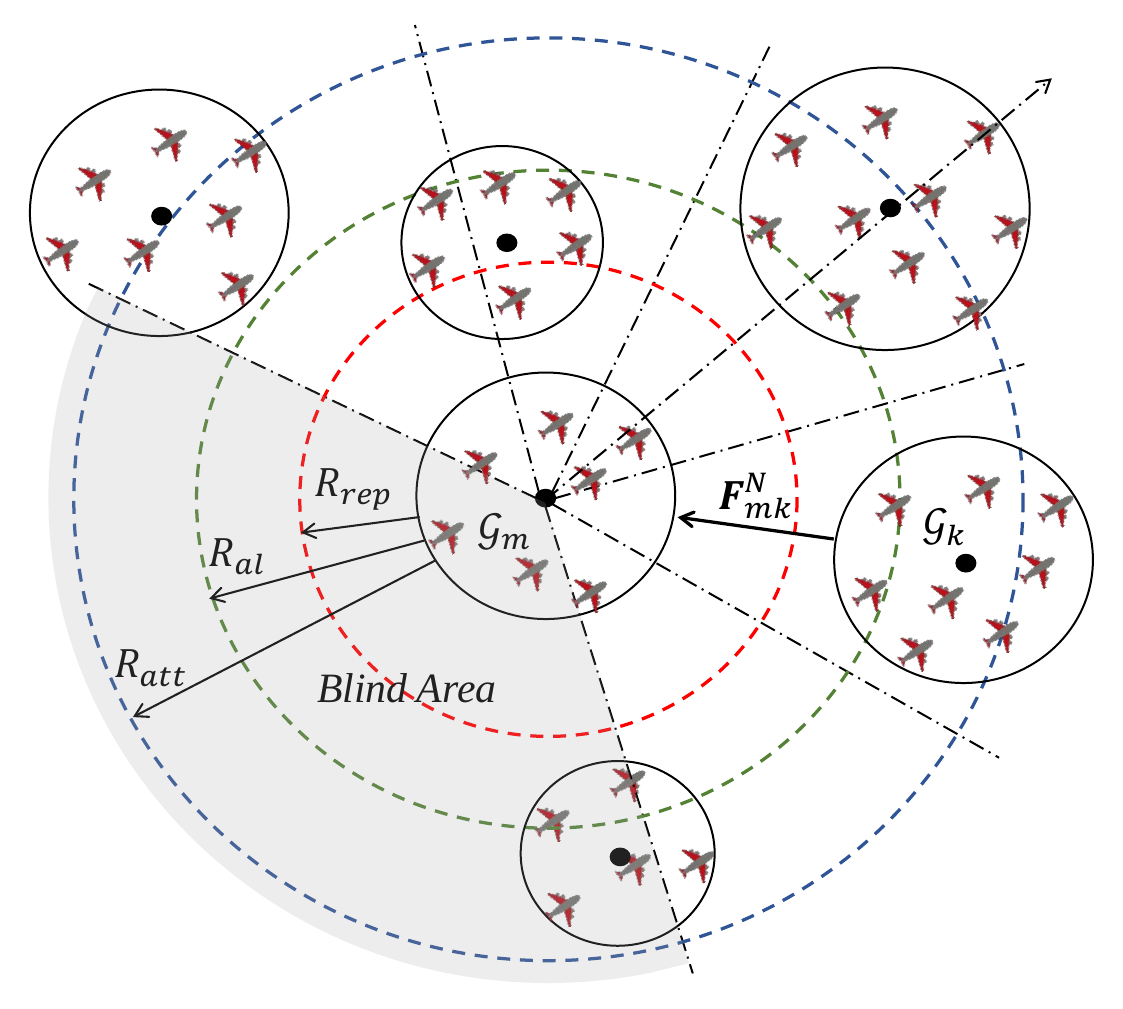}}
\captionsetup{justification=raggedright, singlelinecheck=false}
\caption{Neighboring interaction model of starling-like control with 5 observation regions.}
\label{starling-alg}
\end{figure}

\subsubsection{Neighboring-interaction force}
In the starling formation control, individual starlings only interact with the nearest individuals to avoid too many or too few neighbors in a certain region. Starlings divide the surrounding space into eight regions with an angle of $45^\circ$ each, and only select one neighbor in each region \cite{starling}. To make the starling model more applicable to directional flight scenarios, we form a blind area that covers three regions behind the formation. As shown in Fig.\ref{starling-alg}, the formation $\mathcal{G}_m$ has 5 observation regions to establish the neighboring interaction group set $\mathcal{G}_m^N$. Hence, for the group $\mathcal{G}_k$, we have $\mathcal{G}_k\in\mathcal{G}_m^N$ if and only if $\mathcal{G}_k$ is the nearest formation of $\mathcal{G}_m$ within one of the observation regions. In particular, we have $|\mathcal{G}_m^N|\leq5$.

Moreover, to distinguish between attraction, alignment, and repulsion regions, we also set the attraction radius $R_{att}$, alignment radius $R_{al}$, and repulsion radius $R_{rep}$ for the formations. The interaction force of $\mathcal{G}_m$ caused by $\mathcal{G}_k$ is calculated by (\ref{sneighbor}), 
\begin{table*}[!b]
\normalsize
\centering
\rule{\textwidth}{1pt}
\begin{equation}
\bm{F}_{mk}^N=
\begin{cases} 
    -\dfrac{\bm{D}_{mk}}{\|\bm{D}_{mk}\|}\ln\dfrac{\|\bm{D}_{mk}\|-R_m-R_k}{R_{rep}}, & 0<\|\bm{D}_{mk}\|-R_m-R_k\leq R_{rep}. \\
    0, & R_{rep}<\|\bm{D}_{mk}\|-R_m-R_k\leq R_{al}. \\
    \dfrac{\bm{D}_{mk}}{\|\bm{D}_{mk}\|}\ln\dfrac{R_{att}-(\|\bm{D}_{mk}\|-R_m-R_k)}{R_{att}-R_{al}}, & R_{al}<\|\bm{D}_{mk}\|-R_m-R_k\leq R_{att}.
\end{cases}
\label{sneighbor}
\end{equation}
\end{table*}
where $\bm{D}_{mk}=\bm{P}_k-\bm{P}_m$ is the location difference between $\mathcal{G}_m$ and $\mathcal{G}_k$, $R_m$ and $R_k$ are the equivalent radii of $\mathcal{G}_m$ and $\mathcal{G}_k$ by (\ref{radius}), respectively. Here we use the logarithm function instead of the cosine function in \ref{pneighbor}, which provides smoother attraction and repulsion forces for neighboring interactions. The resultant interactive force applied on formation group $\mathcal{G}_m$ is represented as
\begin{align}
\bm{F}_{m}^N=\mathop{\sum}\limits_{\mathcal{G}_{k}\in\mathcal{G}_{m}^N}\bm{F}_{mk}^N.
\end{align}


\subsubsection{Pattern force}
Unlike the pigeon flocks, where flight information is transmitted through the social hierarchy, the starling flocks exchange flight information in a distributed approach. The starling-like control algorithm generally establishes three motion patterns —collective, evasion, and local following — to achieve collective velocity alignment, obstacle avoidance, and high maneuverability while accounting for the effects of discrete-time communication.

The \textit{collective pattern} is the normal flight model of the starling flocks, which aims to maintain the coordinated flight of the starling flock via velocity balancing. Therefore, the collective pattern force $\bm{F}_m^{Pc}$ is caused by the velocity differences between formation $\mathcal{G}_m$ and its neighboring groups, i.e., 
\begin{align}
    \bm{F}_m^{Pc}=\frac{\mathop{\sum}\limits_{\mathcal{G}_{k}\in\mathcal{G}_{m}^N}(\bm{V}_k-\bm{V}_m)}{\|\mathop{\sum}\limits_{\mathcal{G}_{k}\in\mathcal{G}_{m}^N}(\bm{V}_k-\bm{V}_m)\|},
\end{align}
where $\bm{V}_m$ and $\bm{V}_k$ are the velocities of the leader UAVs of formation group $\mathcal{G}_m$ and $\mathcal{G}_k$ according to (\ref{e2}).

The individuals in starling flocks enter the \textit{evasion pattern} when they detect dangers or obstacles. To simplify the problem description, we assume that each obstacle covers a circular region and the formation performs avoidance maneuvers upon approaching an obstacle within a certain distance. Denote $R_{obs}$ as the obstacle avoidance radius, for each obstacle located at $\bm{P}_{obs}$, the evasion pattern force $\bm{F}_m^{Pe}$ of formation $\mathcal{G}_m$ is calculated as
\begin{align}
    \bm{F}_{m}^{Pe}=
\begin{cases} 
    -\frac{\bm{D}_{mo}}{\|\bm{D}_{mo}\|}\ln\frac{\|\bm{D}_{mo}\|-R_m}{R_{obs}}, &\|\bm{D}_{mo}\|\!-\!R_m\leq R_{obs}. \\
    0, &\|\bm{D}_{mo}\|\!-\!R_m > R_{obs},
\end{cases}
\end{align}
where $\bm{D}_{mo}=\bm{P}_{obs}-\bm{P}_{m}$ is the position difference between the obstacle and the formation.

For nodes that cannot directly detect obstacles, they enter the \textit{local following pattern} when significant velocity changes of their neighbors are observed for obstacle avoidance in advance. For the BINC scheme, the leader node records the velocity of each neighboring group at the previous moment $\bm{V}_k(t-\Delta t)$ and compares it with the current velocity $\bm{V}_k(t)$. If the velocity of a neighboring group changes too quickly, the leader node will replicate the velocity of the neighboring group and perform early obstacle avoidance. Therefore, we need to determine a threshold for pattern transform. Denote $A_{mk}(t)$ as the velocity change indicator of neighboring group $\mathcal{G}_k$ at time $t$, it is calculated as
\begin{align}
    A_{mk}(t) =\frac{1}{\|\bm{D}_{mk}\|}\left(
    1 - \frac{\bm{V}_k(t)}{\|\bm{V}_k(t)\|} \cdot \frac{\bm{V}_k(t-\Delta t)}{\|\bm{V}_k(t-\Delta t)\|}\right).
\end{align}
The neighboring group with the largest velocity change is selected by
\begin{align}
    k^*(t)=\text{arc}\mathop{\text{max}}\limits_{\mathcal{G_k}\in\mathcal{G}_m^N}A_{mk}(t),
    \label{kstar}
\end{align}
and the local-following pattern force $\bm{F}_m^{Pf}$ of formation $\mathcal{G}_m$ is generated as
\begin{align}
    \bm{F}_m^{Pf}=\bm{V}_{k^*}-\bm{V}_{m}.
\end{align}

Then we form an adaptive threshold to evaluate the magnitude of velocity change and determine the pattern transform. A high threshold is required when the velocity consistency of neighboring groups is poor, while we can set a lower threshold if the velocities of neighboring groups are already aligned. Therefore, the threshold of formation $\mathcal{G}_m$ is represented as
\begin{align}
    T_m=\text{exp}\left({-\frac{\alpha}{|\mathcal{G}_m^N|+1}\left\| \frac{\bm{V}_m}{\|\bm{V}_m\|}+\mathop{\sum}\limits_{\mathcal{G}_k\in\mathcal{G}_m^N}{\frac{\bm{V}_k}{\|\bm{V}_k\|}}\right\|}\right),
    \label{alpha}
\end{align}
where $\alpha$ is a positive coefficient to determine the sensitivity of the threshold. With the largest velocity change in (\ref{kstar}), the final pattern force $\bm{F}_m^{P}$ of formation $\mathcal{G}_m$ is represented as
\begin{align}
    \bm{F}_m^{P}=
    \begin{cases}
        \bm{F}_m^{Pe}, & \text{if}\ \|\bm{D}_{mo}\|-R_m\leq R_{obs}.\\
        \bm{F}_m^{Pf}, & \text{if}\ A_{mk^*}> T_m.\\
        \bm{F}_m^{Pc}, & \text{otherwise}.
    \end{cases}
\end{align}

\subsubsection{Velocity loop suppression}
The starlings follow their neighbors continuously in the \textit{local following pattern}; however, in our design, the velocity of groups is periodically broadcast through routing packets. It may take several HELLO cycles to transmit from one group to another group. At this time, when the leader node follows the velocity of its neighbor, it cannot distinguish whether the velocity information is duplicated, resulting in repeated propagation of the information and leading to a Velocity loop issue.

To solve this problem, we add the velocity identification sequence number to the HELLO and C-HELLO messages, as mentioned in Fig.\ref{hello} and Fig.\ref{chello}. The identification sequence number is incremented every time a new velocity is sent. When the leader node decides to follow the group, it will add the ID of the following group and the action identification sequence number to the HELLO message and send it to the follower nodes. When follower nodes learn the velocity from the upper-level nodes, they add the ID of the following group and the action identification sequence number to the HELLO message. After receiving the HELLO message from the upper-level node, the head node records the ID of the following group and the action identification sequence number in the inter-groups C-HELLO message. In this way, when the following chain is formed, the leader node can determine which group the following velocity comes from and can determine whether it has followed that velocity based on the sequence number.

Moreover, to avoid collisions between groups, the frequency of exchanging C-HELLO messages between groups should follow the constraint
\begin{equation}
    T_{C\!-\!HELLO}\times v_{max}<\frac{R_{rep}}{2},
\end{equation}
where $v_{max}$ is the maximum value of the formation flight velocity. In other words, the minimum requirement for two groups to avoid collision in one message cycle is that the motion error does not exceed half of the $R_{rep}$.

So far, we have analyzed the velocity loop issue in discrete-time scenarios and proposed corresponding mechanisms to address this problem. We established the mapping from the starling local-following pattern to group flight in discrete-time communication scenarios, enhancing the maneuvering performance of the BINC scheme with small communication overheads.

\section{Simulation Results}
This section first introduces the setup of simulation experiments. Then, performance analysis on communication overhead, network maintenance, and control maneuverability of the proposed scheme and other approaches are presented in the rest subsections.

\subsection{Simulation Setup}
Qualnet is a network simulation software developed by Scalable Networks Technologies. All UAVs are randomly initialized within a $60000\times60000 m^2$ square area with zero initial velocities. The number of UAVs in the swarm is varying as $N\in[520,640,760,880,1000]$. To better analyze the control performance of different approaches, we form two flight scenarios - straight sailing and obstacle avoidance. For the straight sailing scenario, the destination is set to a distant point on the right side of the swarm, hence the UAV swarm will fly straight to the right. For the obstacle avoidance scenario, we established a circle obstacle located at the middle of the swarm flight path with $\bm{P}_{obs}=[14000m,0m]^\top$ and $R_{obs}=5400m$, forcing the UAV swarm to circumvent. Other parameters of the simulation are configured in Table.\ref{parameters}.
\begin{table}[ht]
    \centering
    \caption{Simulation Parameters Setup}
    \begin{tabular}{ccc}
    \hline
    Parameter & Default Value & Description\\
    \hline\hline
    $v_{max}$ & 20 m/s & Maximum flight velocity \\
    $r_{rep}$  & 100 m & Pigeon-like repulsive radius\\
    $r_{al}$  & 150 m & Pigeon-like align radius \\
    $r_{att}$  & 200 m & Pigeon-like attractive radius\\
    $R_{rep}$  & 200 m & Starling-like repulsive radius \\
    $R_{al}$  & 400 m & Starling-like align radius \\
    $R_{att}$  & 600 m & Starling-like attractive radius \\
    $d_{tr}$ & 200 m & Short-channel communication distance \\
    $D_{tr}$ & 1000 m & Long-channel communication distance \\
    $A$ & 0.5 & Coefficient $A$ in (\ref{pneighbor})\\
    $\alpha$ & 1 & Coefficient $\alpha$ in (\ref{alpha})\\
    $T_{HELLO}$ & 2 s & HELLO message interval\\
    $T_{C\!-\!HELLO}$ & 2 s & C-HELLO message interval\\
    $T_{TC}$ & 5 s & TC message interval\\
    $T_{HTC}$ & 5 s & HTC message interval\\
    $T_{CMN}$ & 2 s & CMN message interval\\
    $T_{MWT}$ & 10 s & MAX\textunderscore WAITING\textunderscore TIME\\
    \hline
    \end{tabular}
    \label{parameters}
\end{table}

\subsection{Analysis on Communication Overhead}
\begin{figure}[!t]
\centerline{\includegraphics[width=.96\linewidth]{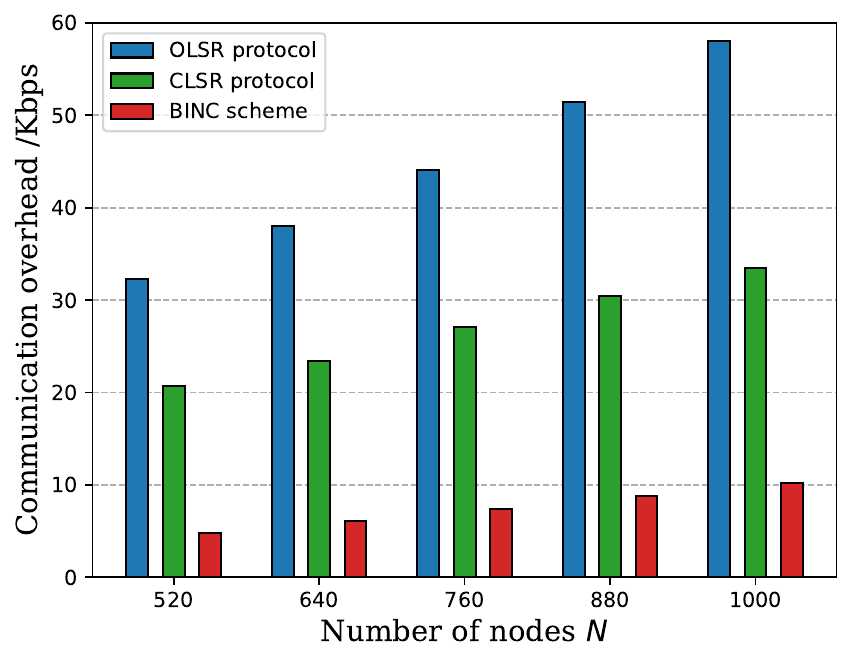}}
\caption{Results of average communication overhead with different protocols under different numbers of UAVs $N$.}
\label{overhead1}
\end{figure}

We count the communication overheads of the BINC scheme under the straight sailing scenario for swarms with 520, 640, 760, 880, and 1000 nodes, and compare the results with OLSR \cite{olsr} and CLSR \cite{clsr} protocols. Each simulation is randomly repeated 10 times. As shown in Fig.\ref{overhead1}, the communication overhead of all approaches increases linearly with the increase of the number of nodes.
Meanwhile, the proposed BINC scheme requires the minimum overheads to achieve large-scale swarm networking and control, reducing the routing overhead by an average of 85.2\% compared to OLSR and 70.5\% compared to CLSR. Therefore, the total cost of the BINC scheme can meet the routing and communication requirements of large-scale UAV swarms.

\begin{figure}[!t]
\centerline{\includegraphics[width=.96\linewidth]{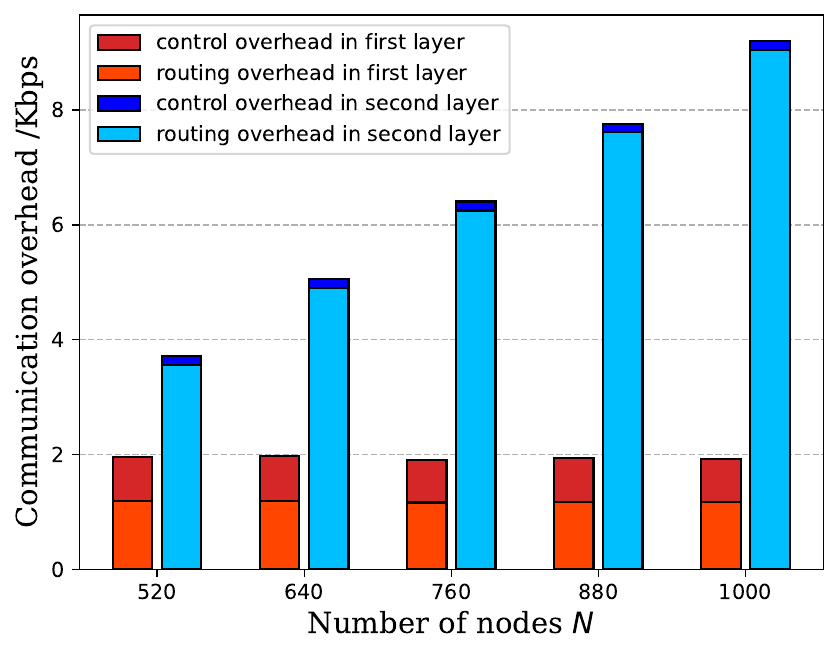}}
\caption{Results of average overhead caused by routing and control within different layers under different numbers of UAVs $N$.}
\label{overhead2}
\end{figure}

To evaluate the effectiveness of the hierarchical structure and the fused message design, we further analyze the communication overheads caused by routing and control within different layers of the proposed BINC scheme. Similarly, we simulate the straight sailing scenario for swarms with 520, 640, 760, 880, and 1000 nodes 10 times. As shown in Fig.\ref{overhead2}, the communication overheads within the first layer remain at around 2 Kbps. This happens as the size of each cluster is determined by the $2$-hop range. Therefore, the increasing scale of the swarm only leads to an increase in the number of clusters, but won't affect the size of each cluster much. The communication overheads within the second layer increase linearly as the number of UAVs increases, but this overhead is limited to less than 10 Kbps even below a scale of 1000 nodes.


The results in Fig.\ref{overhead2} also demonstrate the proportion of overheads caused by routing and control requirements. For the first layer, the routing overhead is around 1.2 Kbps. For the second layer, as the HTC messages are broadcast to the entire network, the routing overhead increases according to the number of clusters. Meanwhile, the control overhead within the first layer is about 0.8 Kbps due to the CMN messages in the pigeon-inspired algorithm. On the contrary, the second layer requires only about 0.15 Kbps of control cost, since each formation in the starling-inspired algorithm only cares about no more than 5 neighboring groups for flight information transfer. In summary, the number of UAVs only affects the routing overhead of the second layer, hence, the proposed BINC scheme enables robust scalability for large-scale swarm control with low overhead.

\subsection{Analysis on Network Maintenance}
\begin{figure}
    \centering
    \includegraphics[width=.96\linewidth]{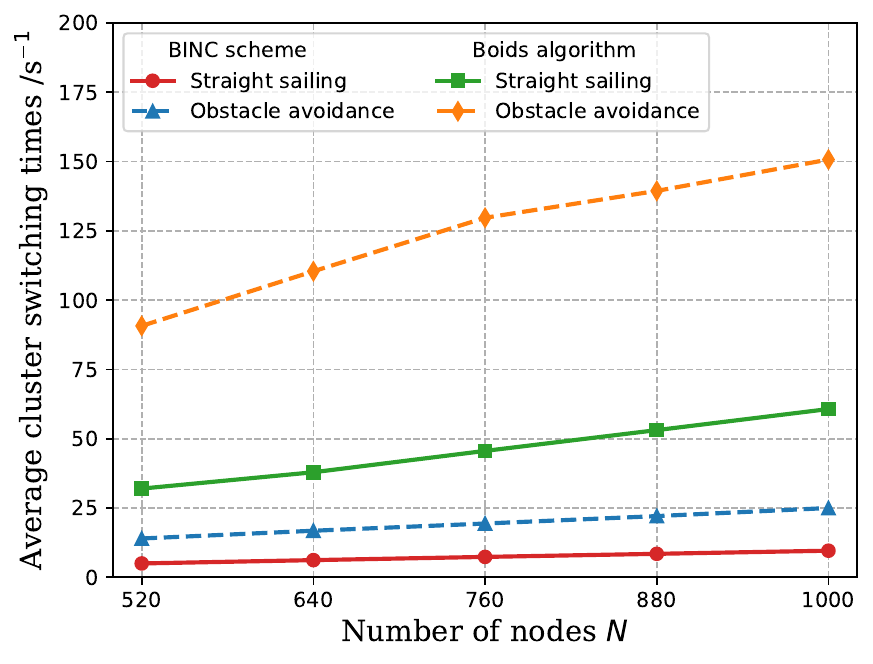}
    \caption{Results of the number of cluster switches with different control algorithms under different numbers of UAVs $N$.}
    \label{maintain1}
\end{figure}

The important innovation of the proposed scheme is the combination of formation control and network maintenance. The BINC scheme prevents the mutual influence of different network clusters while maintaining multi-group flight. We use average cluster switching times in different clusters to characterize the network maintenance capability of the BINC scheme. We compared the Boids algorithm with the BINC scheme and conducted simulation experiments in straight sailing scenarios and obstacle avoidance scenarios with 520, 640, 760, 880, and 1000 UAVs, respectively. As plotted in Fig.\ref{maintain1}, the switch of clusters happens more frequently under the obstacle avoidance scenarios, due to dramatic velocity changes of the formations near the obstacle. Moreover, the results prove that the BINC scheme can effectively reduce the number of clustering switches compared to the Boids algorithm in varying scenarios, hence achieving more robust network topologies. 
\begin{figure}[!t]
\centerline{\includegraphics[width=.96\linewidth]{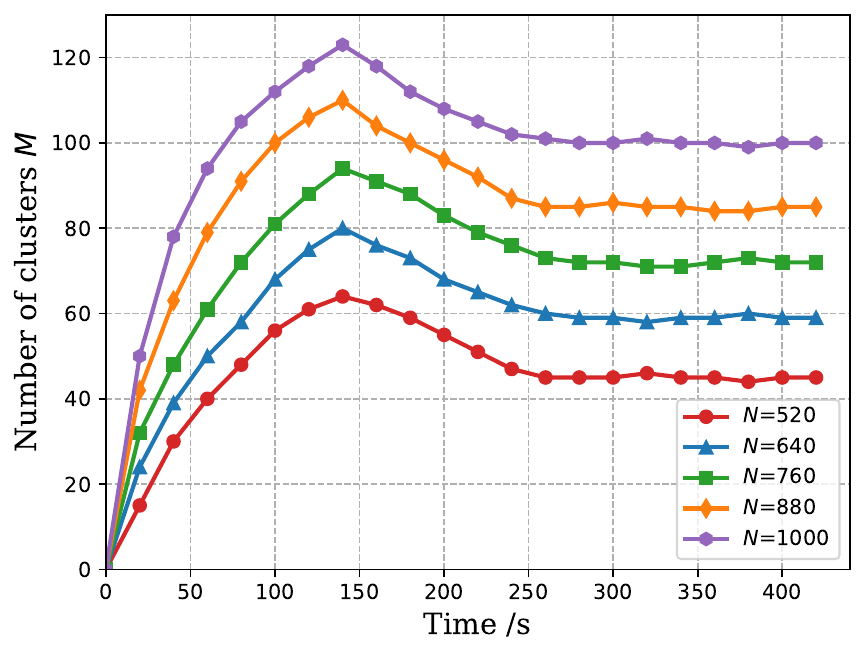}}
\captionsetup{justification=raggedright, singlelinecheck=false}
\caption{Results of the clustering process of the BINC scheme under different numbers of UAVs $N$.}
\label{maintain2}
\end{figure}

To observe the clustering process of the proposed scheme, we illustrate the number of clusters under straight sailing scenarios with different $N$ in Fig.\ref{maintain2}. It can be seen that clustering can be divided into two stages. In the first stage, the number of clusters steadily increases and reaches its peak. This is because after clustering begins, the node with the highest connective degree within the two-hop range becomes the head node and forms a cluster. In the second stage, the number of clusters decreases and eventually reaches a stable state caused by the fusion of clusters. In addition, it can be seen that there is a positive correlation between the number of clusters and the size of the swarm; the larger the cluster size, the more clusters there are.

\begin{figure}[!t]
\centerline{\includegraphics[width=.98\linewidth]{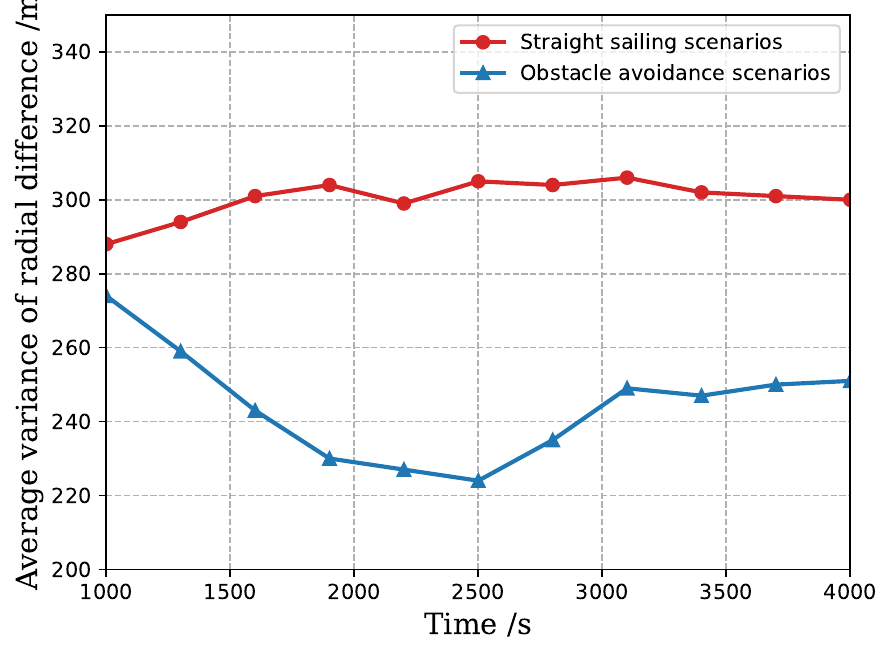}}
\captionsetup{justification=raggedright, singlelinecheck=false}
\caption{Results of average radial difference of the BINC scheme under different scenarios.}
\label{maintain3}
\end{figure}

To evaluate the control effect of the BINC scheme on inter-formation distance, we plot the variation of average radial difference in both straight sailing and obstacle avoidance scenarios with 520 nodes. As shown in Figure \ref{maintain3}, the results demonstrate that the BINC scheme can maintain the radial difference within a stable range in both scenarios, without experiencing drastic fluctuations over time. For obstacle avoidance scenarios, due to the obstruction effect of obstacles, the average radial difference will tend to be closer to the size of the exclusion radius, making the formation dense.

\begin{figure}[!t]
\centerline{\includegraphics[width=.96\linewidth]{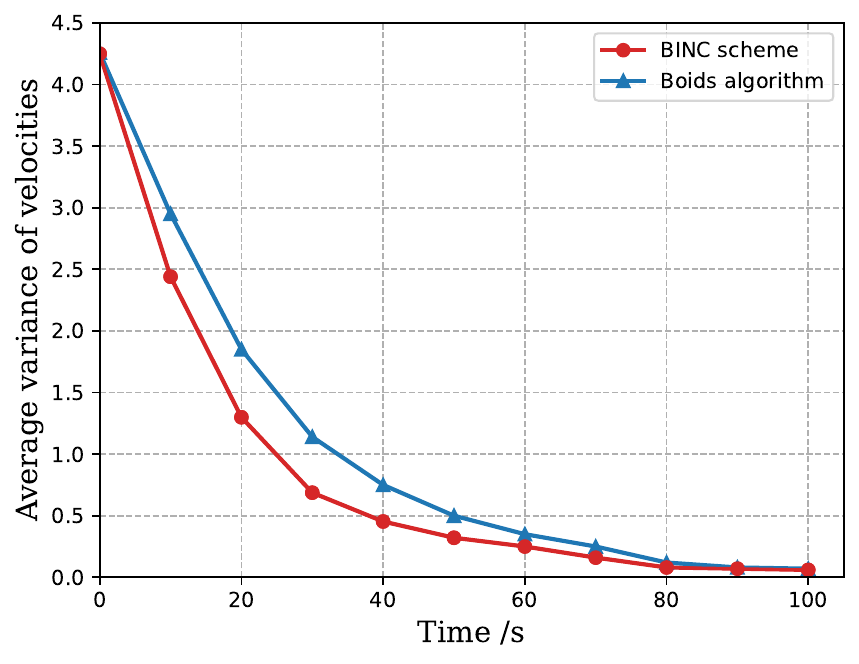}}
\captionsetup{justification=raggedright, singlelinecheck=false}
\caption{Results of average variance of velocity with different control algorithms.}
\label{control1}
\end{figure}

\subsection{Analysis on Swarm Control}

Since the control algorithms aim to enforce velocity consensus across neighboring agents within UAV swarms, we conduct a comparative analysis focusing on velocity convergence time and alignment precision. The evaluation leverages a swarm of 520 UAVs initialized with randomized velocity vectors. Velocity polarization efficacy is quantified via the average variance of node velocity vectors. The temporal evolution of velocity variance for both algorithms, illustrated in Fig. \ref{control1}, reveals that the BINC scheme achieves marginally faster convergence while maintaining alignment accuracy comparable to the Boids algorithm, with a 21.6\% reduction of velocity variance within the first minutes. 
This study underscores BINC’s scalability advantage in large-scale deployments, where hierarchical coordination mitigates latency penalties associated with fully distributed consensus protocols.

\begin{figure*}[!t]
\centering{
\subfloat[trajectories of the BINC scheme]{\includegraphics[width=.322\linewidth]{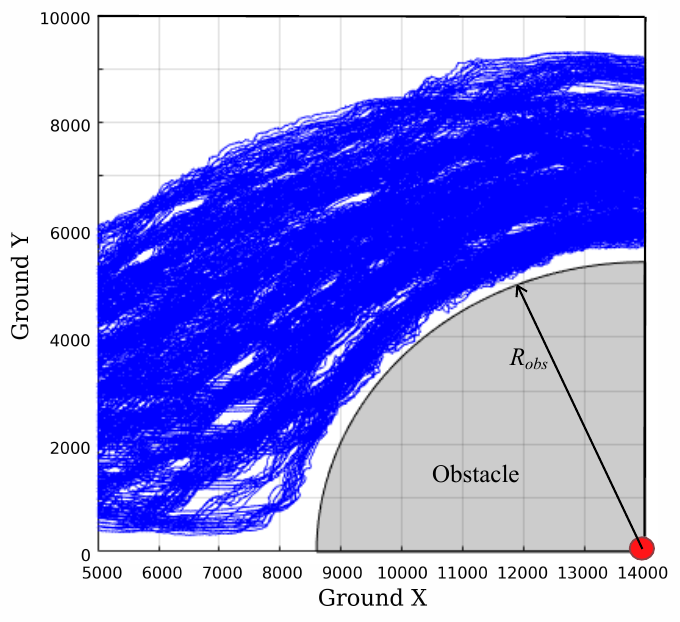}\label{control2a}}
\subfloat[trajectories of the Boids algorithm]{\includegraphics[width=.322\linewidth]{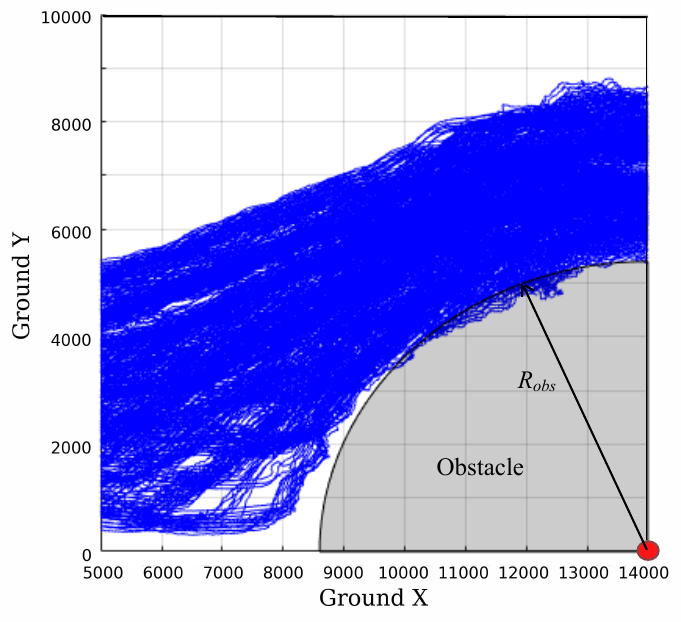}\label{control2b}}
\subfloat[minimum distance to the obstacle]{\includegraphics[width=.34\linewidth]{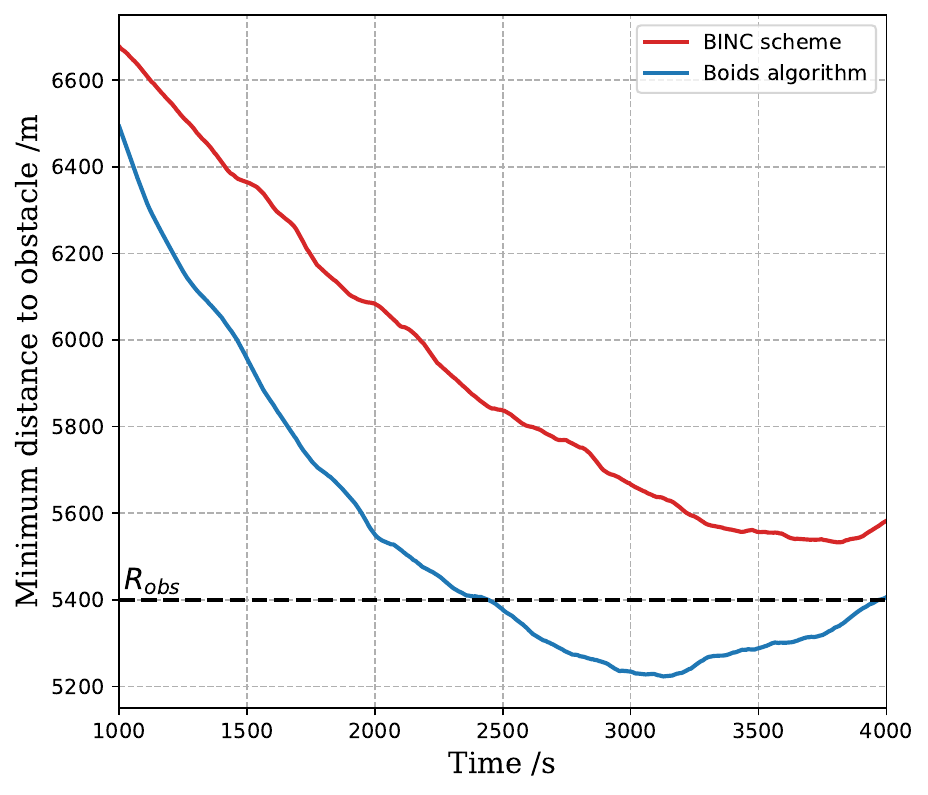}\label{control2c}}}
\captionsetup{justification=raggedright, singlelinecheck=false}
\caption{Results of the obstacle avoidance trajectories and minimum distances between obstacle and swarm with different control algorithms.}
\label{control2}
\end{figure*}


Another significant innovation of the BINC scheme is its integration of a starling-inspired collision avoidance mechanism, which enhances swarm survivability through proactive obstacle evasion. To validate this capability, we performed large-scale obstacle avoidance experiments involving 520 UAVs navigating toward a target square along the x-axis in a Cartesian coordinate system. The obstacle avoidance trajectories of both control algorithms are plotted in Fig.\ref{control2}a-b, respectively. Comparative analysis reveals distinct behavioral patterns: the BINC scheme exhibits a broader turning radius during obstacle evasion, enabling earlier course adjustments. The trajectories controlled by the BINC scheme in Fig.\ref{control2a} fully eliminate boundary incursions, while the trajectories of the Boids algorithm in Fig.\ref{control2b} demonstrate partial UAV trajectories intersecting the obstacle boundary during high-density maneuvers. To quantitatively evaluate the obstacle avoidance effect of the BINC scheme, we compared the minimum distance between the UAVs and the obstacle center of different approaches. As shown in Fig. \ref{control2c}, when obstacles approach, the starling obstacle avoidance mechanism keeps the distance larger than the radius of the obstacle. The minimum distance to the obstacle of the BINC scheme is 5565m, outperforming the Boids algorithm with a 6.36\% improvement in obstacle avoidance radius. Therefore, the BINC scheme achieves highly maneuverable turning and obstacle avoidance for swarm control.

\section{Conclusion}
In this paper, we study the dual challenges of scalable networking and coherent formation control in large-scale UAV swarms, and introduce a bio-inspired hierarchical architecture that synergistically integrates cluster-based routing with bio-inspired control strategies. The proposed BINC scheme establishes a two-tiered topology through localized two-hop neighbor clustering, effectively constraining routing overhead by limiting broadcast domains. Building upon this partitioned structure, we implement distinct bio-inspired control paradigms: intra-cluster cohesion governed by pigeon-inspired leader-follower interactions and inter-cluster alignment guided by starling-like collision avoidance mechanisms. Simulation has proven that the BINC scheme enables efficient network routing and maneuverable formation controlling in large-scale UAV swarms with up to 12 Kbps of communication overhead.



\bibliographystyle{IEEEtran}
\balance
\bibliography{IEEEabrv,main}


 




\vfill

\end{document}